\documentclass[a4paper,11pt]{article}
\usepackage{pos}

\usepackage{caption}
\usepackage{subcaption}
\newcommand{\eq}[1]{\eqref{eq:#1}}
\newcommand{\fig}[1]{Fig.~\ref{fig:#1}}
\newcommand{\tab}[1]{Tab.~\ref{tab:#1}}

\title{Old and new anomalies in charm}

\author[a]{Rigo Bause}
\author[b]{Hector Gisbert}
\author[a]{Gudrun Hiller}
\author*[a]{Tim Höhne}
\author[c]{Daniel F. Litim}
\author[a, d]{Tom Steudtner}

\affiliation[a]{Department of Physics, TU Dortmund University,\\
 Otto-Hahn-Str. 4, D-44221 Dortmund, Germany}

\affiliation[b]{Istituto Nazionale di Fisica Nucleare (INFN), Sezione di Padova, \\
Via F. Marzolo 8, 35131 Padova, Italy \\[1mm]
Dipartimento di Fisica e Astronomia `G.~Galilei', Universit\`a di Padova,\\ 
Via F. Marzolo 8, 35131 Padova, Italy \\[1mm]}

\affiliation[c]{Department of Physics and Astronomy, University of Sussex,\\
 Brighton, BN1 9QH, U.K.}

\affiliation[d]{Department of Physics, University of Cincinnati,\\
 Cincinnati, OH 45221, USA}

\emailAdd{rigo.bause@tu-dortmund.de}
\emailAdd{hector.gisbert@pd.infn.it}
\emailAdd{gudrun.hiller@tu-dortmund.de}
\emailAdd{tim.hoehne@tu-dortmund.de}
\emailAdd{d.litim@sussex.ac.uk}
\emailAdd{tom2.steudter@tu-dortmund.de}

\abstract{The recent LHCb determination of the direct \textit{CP} asymmetries in the decays $D^0 \to K^+ K^-, \pi^+ \pi^-$ hints at a sizeable breaking of two approximate symmetries of the SM: \textit{CP} and U-spin.
We aim at explaining the data with BSM physics and use the framework of flavorful $Z^\prime$ models. Interestingly, experimental and theoretical constraints very much narrow down the shape of viable models: Viable, anomaly-free models are electron- and muon-phobic and feature a light $Z^\prime$ of 10-20 GeV coupling only to right-handed fermions.
The $Z^\prime$ can be searched for in low mass dijets or at the LHC as well as dark photon searches.
A light $Z^\prime$ of $\sim$ 3 GeV or $\sim$ 5-7 GeV can moreover resolve the longstanding discrepancy in the $J/\psi, \psi^\prime$ branching ratios with pion form factors from fits to $e^+ e^- \to \pi^+ \pi^-$ data, and simultaneously explain the charm \textit{CP} asymmetries. 
Smoking gun signatures for this scenario are $\Upsilon$ and charmonium decays into pions, taus or invisbles.
}

\FullConference{21st International Conference on B-Physics at Frontier Machines (Beauty2023)\\
 3-7 July, 2023\\
Clermont-Ferrand, France\\}

\begin{document}

\renewcommand{\hookAfterAbstract}{%
\par\bigskip
\textsc{ArXiv ePrint}:
\\Report number: DO-TH 23/14
}

\maketitle

\section{Introduction}
\vspace{-.2cm}

\begin{figure}[h!]
  \begin{minipage}{0.57\textwidth}
    \centering
    \includegraphics[width=0.9\linewidth]{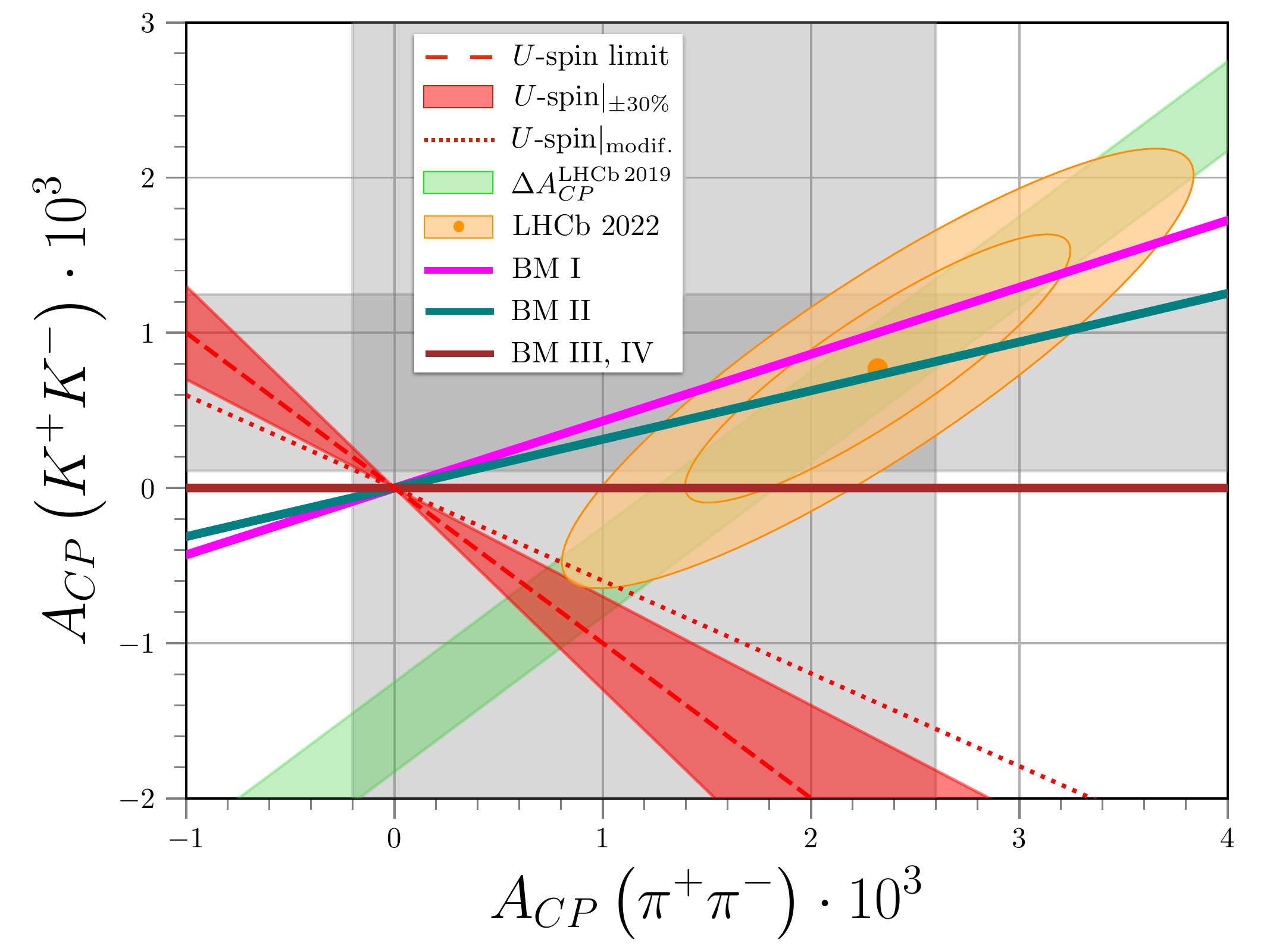}
    \caption{The U-spin-\textit{CP} anomaly in charm, showing LHCb bounds (gray \& green), the best fit region (orange) and the SM expectation including $\lesssim 30\%$ U-spin breaking (red cone). Coloured lines relate to benchmark models of \cite{Bause:2022jes}.}
    \label{fig:Anomaly}
  \end{minipage}
  \hfill
  \begin{minipage}{0.37\textwidth}
    \hspace{-.5cm}
    \includegraphics[width=1.1\linewidth]{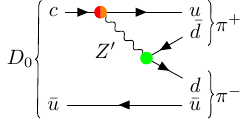}
    \\ \vspace{0.7cm} \\
    \caption{$Z^\prime$ contribution to the decay $D^0 \to \pi^+ \pi^-$.}
    \label{fig:Dtopipi}
  \end{minipage}
\end{figure}

Recently, LHCb determined the direct \textit{CP} asymmetries in $D^0 \to \pi^+\pi^-,K^+K^-$ decays 
to \cite{LHCb:2022vcc} \vspace{-.3cm}
\begin{equation} \label{eq:ad}
a_{K^+K^-}^d = (7.7 \pm 5.7) \cdot 10^{-4}, \qquad
a_{\pi^+ \pi^-}^d = (23.2 \pm 6.1) \cdot 10^{-4}.
\end{equation} \vspace{-.9cm}\\
These results are puzzling for two reasons. 
Firstly, a SM explanation of $a_{\pi^+ \pi^-}^d$ requires higher-order contributions $h$ to be enhanced over the tree-level amplitude $t$ by $\tfrac{h}{t}\sim 2$. This is significantly larger than the estimations from \cite{Khodjamirian:2017zdu,Pich:2023kim}.
Secondly, the fit implies a $2.7 \sigma$ violation of the approximate SM U-spin symmetry \cite{Schacht:2022kuj}, that is $a_{\pi^+\pi^-}^d\!=\!-a_{K^+K^-}^d$, which is a factor $\sim\!4$ larger than the naively expected U-spin breaking in the SM of $\mathcal{O}\left(\frac{m_s-m_d}{\Lambda_{QCD}}\right) \sim 30\%$, see \fig{Anomaly}.
This constitutes the U-spin-\textit{CP} anomaly in charm, which we aim to explain with a flavorful $Z^\prime$ model \cite{Bause:2022jes}.

\vspace{-.1cm}

\vspace{-0.5cm}

\section{Explaining the anomaly with a flavorful $Z^\prime$}
\vspace{-.2cm}

The $Z^\prime$ contributes to the \textit{CP} asymmetries in $D^0 \to \pi^+\pi^-,K^+K^-$ decays via \fig{Dtopipi} as
\vspace{-.3cm}
\begin{equation}
a_{\pi^+\pi^-,K^+K^-}^d = \frac{g_4^2}{M^2_{Z^\prime}}\Delta \widetilde{F}_R \left[c_{\pi,K} F_{Q_{1,2}} + d_{\pi,K}F_{d_{1,2}}\right]
\end{equation} \vspace{-.7cm}\\
where $g_4$ and $M_Z^\prime$ are the $U(1)^\prime$ gauge coupling and $Z^\prime$ mass, respectively, and $c_{\pi,K}, d_{\pi,K}$ are hadronic parameters. Moreover, $\Delta \widetilde{F}_R= \sin \theta_u \cos \theta_u (F_{u_2}-F_{u_1})$ contains the right-handed $c$-$u$ mixing angle $\theta_u$.
Explaining the \textit{CP} data \eq{ad} requires the $U(1)^\prime$-quark charges to obey $F_{u_2}\neq F_{u_1}$ and $\vert F_{d_1} \vert \gg \vert F_{d_2} \vert$ due to the hierarchy $a^d_{\pi^+\pi^-} \gg a^d_{K^+K^-}$ \eq{ad}, along with $\theta_u \neq 0$ and sizeable relative weak and strong phases.

The shape of viable benchmark models in \tab{BMs} is further narrowed down by demanding anomaly cancellation which might require adding $U(1)^\prime$ charged right-handed neutrinos $\nu_R$. Additional constraints arise from Kaon FCNCs, (semi-)leptonic and (semi-)invisible $D \to (\pi) \ell^+ \ell^-,\nu \nu$ decays as well as Drell-Yan searches.
Viable models also predict $A_{CP}(\pi^0\pi^+)\! \simeq \!A_{CP}(\pi^0\pi^0)\! \simeq \! + 10^{-3}$.

Moreover, strong constraints from $D$-mixing combined with \textit{CP} data \eq{ad} surpisingly point to a sub-electroweak $Z^\prime$ mass of a few $\times 10$ GeV.
The $Z^\prime$ coupling to $d$-quarks leads to collider signals in low mass dijets 
with initial state radiation, resulting in a mass bound of $M_{Z^\prime} \lesssim 20$ GeV \cite{CMS:2019xai}.

\clearpage


\begin{table}[t]
\resizebox{1\textwidth}{!}{
\begin{tabular}{c|rcr|rcr|rcr|rcr|rcr|rcr|rcr}
\hline \hline
  Model &&$F_{Q_i}$&&& $F_{u_i}$ &&& $F_{d_i}$ &&& $F_{L_i}$ &&& $F_{e_i}$ &&& $F_{\nu_i}$ & \\ \hline \hline
  BM III & 0 & 0 & 0 &   $0\quad$ & -1 & 0 & 1 & 0 & 0 & 0 & 0 & 0 & $0\;\;$ & 0 & $1\;\;$ & 0 & 0 & -1 \\
  BM IV  & 0 & 0 & 0 & -$\tfrac{985}{1393}$ & $\tfrac{985}{1393}$  & 0 & 1 & 0 & -$1$ & 0 & 0 & 0 & $\tfrac{1}{1393}$ & 0 & -$\tfrac{1}{1393}$ & $F_\nu$ & -$F_\nu$ & 0  \\
  \hline \hline
  \end{tabular}}
\caption{Benchmarks for viable, anomaly-free $U(1)^\prime$ extensions of the SM+3$\nu_R$ explaining $a_{\pi^+\pi^-,K^+K^-}^d$ \eq{ad}. \vspace{-.58cm}
}
\label{tab:BMs}
\end{table}

\section{A hadrophilic $Z^\prime$ of $\mathcal{O}$(10 GeV)?}
\vspace{-.2cm}

Light $Z^\prime$ models are also constrained by dark photon searches \cite{LHCb:2019vmc}, which imply a strict bound on lepton charges of $\vert F_{L_{1,2},e_{1,2}} \vert \lesssim 10^{-3} \vert F_{d_1}\vert $. Hence, the $Z^\prime$ has to be leptophobic.

\begin{figure}[h!]
  \begin{minipage}{0.37\textwidth}
    \centering
    \vspace{.3cm}
    \includegraphics[width=1.15\linewidth]{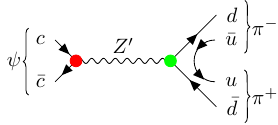}
    \\ \vspace{1.2cm}
    \caption{$Z^\prime$ contribution to the charmonia decays $J/\psi, \psi^\prime \to \pi^+ \pi^-$.}
    \label{fig:Psitopipi}
  \end{minipage}
  \hfill
  \begin{minipage}{0.57\textwidth}
    \centering
    \includegraphics[width=0.9\linewidth]{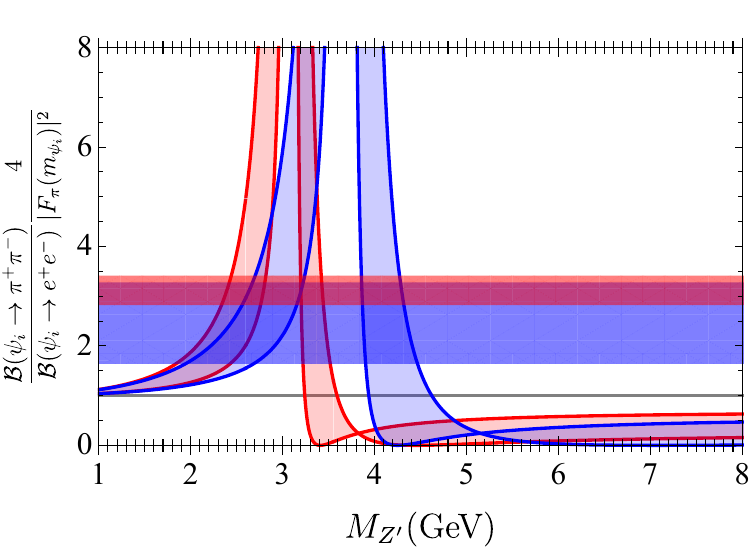}
    \caption{Constraints from charmonium decays. Horizontal red (blue) bands denote $J/\psi$ ($\psi^\prime$) data \cite{Bruch:2004py} using $F_\pi(q^2)$ from \cite{Cheng:2020vwr}. Curves correspond to predictions in BM III including experimental uncertainties from \eq{ad}. The SM prediction corresponds to the horizontal gray line.}
    \label{fig:Brs}
  \end{minipage}
\end{figure}

The $Z^\prime$ can also mediate quarkonia decays. In BM IV, mass bounds arise from $\Upsilon(1s)$ decays via the $b_R$-coupling.
Moreover, the $Z^\prime$ contributes to charmonium decays $\psi_i \to Z^{\prime*} \to \pi^+ \pi^-(,\tau^+ \tau^-, \nu \nu)$ with $\psi_i=J/\psi, \psi^\prime$, see \fig{Psitopipi}.
In particular, the electrophobic $Z^\prime$ enhances $\mathcal{B}(\psi_i \to \pi^+ \pi^-)$ with respect to $\mathcal{B}(\psi_i \to e^+ e^-)$, see \fig{Brs}. Thereby, for $M_{Z'}\simeq 3$ GeV (5-7 GeV) in BM III (BM IV) the model is able to resolve the longstanding discrepancy between the pion form factor $F_\pi(q^2)$ extracted from $J/\psi \to \pi^+ \pi^-$ \cite{Bruch:2004py} and $e^+ e^- \to \pi^+ \pi^-$ \cite{Cheng:2020vwr}.
In this case, in BM III the $Z^\prime$ mass range of $M_{Z^\prime} \lesssim 2.2$ GeV allowed by $\mathcal{B}(\psi^\prime \to \tau^+ \tau^-) = (3.1 \pm 0.4)\cdot 10^{-3}$ \cite{ParticleDataGroup:2022pth} almost coincides with the pion form factor window, whereas \mbox{$\mathcal{B}(J/\psi \to \text{nothing}) < 7\cdot 10^{-4}$} \cite{ParticleDataGroup:2022pth} implies that the decay to BSM neutrinos should be kinematically forbidden by $M_\nu > M_{J/\Psi}/2$ which can be achieved e.g. via the Dirac inverse see-saw mechanism.

\vspace{-.5cm}

\vspace{-0.7cm}

\section{Conclusion}
\vspace{-.2cm}
Recent charm data \eq{ad} imply sizeable violation of \textit{CP} and U-spin, possibly hinting new physics. We obtain a viable explanation from a flavorful $Z^\prime$ boson which is light of $\mathcal{O}$(10 GeV), leptophobic and couples only to $SU(2)_L$ singlets. 
Moreover, a $Z^\prime$ of a few GeV can resolve the pion form factor discrepancy between $e^+e^-\to \pi^+ \pi^-$ and $J/\psi \to \pi^+ \pi^-$ extractions.
In this scenario, hadronic, tauonic and invisible quarkonia decays are smoking gun signatures of the model. 
Longstanding anomalies such as the large isospin breaking between $\psi(3770) \to D^+D^-$ and $D_0 \bar D_0$ \cite{ParticleDataGroup:2022pth} could potentially also be addressed. \vspace{-.45cm} 

\clearpage

\acknowledgments{TH would like to thank the organizers for the invitation to such a stimulating conference. This work is supported by the \textit{Studienstiftung des deutschen Volkes} (TH), the \textit{Bundesministerium für
Bildung und Forschung (BMBF)} under project number 05H21PECL2 (HG), and the Science and Technology Research Council (STFC) under the Consolidated Grant ST/T00102X/1 (DFL).}

\end{document}